\documentclass[
prl,
twocolumn,
showpacs,
preprintnumbers,
amsmath]{revtex4}

\usepackage{graphicx}	

\usepackage{amssymb}	
\usepackage{amsmath}	
\usepackage{mathptm}	
\usepackage{color}	

\newcommand{\be}{\begin{equation}}
\newcommand{\ee}{\end{equation}}
\newcommand{\bea}{\begin{eqnarray}}
\newcommand{\eea}{\end{eqnarray}}

\def\Corr#1{#1}				

\begin{document}       


\preprint{Physical Review Letters, in press (2009)}

\title{Prediction of spatio-temporal patterns of neural \\
activity from pairwise correlations}

\author{O.~Marre, S.~El Boustani, Y.~Fr\'egnac,
A.~Destexhe$^{*}$} 

\affiliation{{\small\sl Unit\'e de Neurosciences Int\'egratives et Computationnelles (UNIC), UPR CNRS 2191, Gif-sur-Yvette, France} \\
$*$ Corresponding Author: destexhe@unic.cnrs-gif.fr\\
}


\date{\today}



\begin{abstract}
We designed a model-based analysis to predict the occurrence of
population patterns in distributed spiking activity. Using a maximum
entropy principle with a Markovian assumption, we obtain a model
that accounts for both spatial and temporal pairwise correlations
among neurons. \Corr{This model is tested on data generated with a Glauber
spin-glass system and is shown to correctly predict the occurrence
probabilities of spatio-temporal patterns significantly better than
Ising models taking into account only pairwise correlations. This
increase of predictability was also observed on experimental data
recorded in parietal cortex during slow-wave sleep.}  This approach
can also be used to generate surrogates that reproduce the spatial
and temporal correlations of a given data set.
\end{abstract}


\maketitle


The structure of the cortical activity, and its relevance to sensory processing or motor planning, are a long standing debate \cite{Abeles1982}. There is a need to describe the structure of the spiking activity based on well-defined statistical models. 
To infer the state of the neural network, a first line of work has tried to model the neural activity with Hidden Markov Models \cite{Radons1994b,Abeles1995,Yu2006}. \Corr{Maximum entropy models have proved useful for the analysis of many complex systems (see for example \cite{Lezon2006,Berger1996})). 
Another line of research has used this approach to describe neural activity, focusing on 
spiking patterns lying within one time bin \cite{Schneidman2006,Shlens2006}. However, the latter is not prone} to predict the temporal statistics of the neural activity \cite{Tang2008}. 
In the following study, we design a model inspired from both lines of research to better describe the neural dynamics. 
This model is a maximum entropy model based on the correlation values, and respecting a Markovian assumption. Thus it takes into account both spatial and temporal correlations. We show its 
ability to describe 
the spatio-temporal statistics of the activity on simple network models \Corr{and recordings in the mammalian parietal cortex {\it in vivo}.} 

We consider $N$ neurons whose spikes are recorded and binned, for a long time period, noted as $\{\sigma(t)\} := \{ \sigma_{i}(t)\}_{i=1,...,N}$ \Corr{where $\sigma_i \in \{-1;1\}$}. The purpose of a statistical model is to describe as \Corr{closely} as possible the probability distribution of the spatio-temporal patterns, $P(\{\sigma(t)\} ,\{\sigma(t+1)\}, ...)$ with a limited number of parameters. For that purpose, we make a Markovian hypothesis on this distribution, and aim at finding the joint distribution $P(\{\sigma\}^{\tau+1};\{\sigma'\}^{\tau})=P(\{\sigma\}^{\tau+1}|\{\sigma'\}^{\tau})P(\{\sigma'\}^{\tau})\label{eq:JointDistrib}$ which maximizes the entropy $H(\{\sigma\}^{\tau+1};\{\sigma'\}^{\tau})=-\sum_{\{\sigma\},\{\sigma'\}}P(\{\sigma\}^{\tau+1};\{\sigma'\}^{\tau})\ln\left(P(\{\sigma\}^{\tau+1};\{\sigma'\}^{\tau})\right)\label{eq:Entropy}$ with the constraints on the first- and second-order statistical moments of the activity $m_{i} = <\sigma_{i}>$, $C_{ij} = <\sigma_{i}(t)\sigma_{j}(t)>$ and $C_{ij}^{1} = <\sigma_{i}(t)\sigma_{j}(t+1)>$, 
the normalisation constraint, and the marginal distribution constraint: $\sum_{\{\sigma'\}}P(\{\sigma\}^{\tau+1};\{\sigma'\}^{\tau})=P(\{\sigma\}^{\tau+1})$.

By using Lagrange multipliers, and then applying the marginal distribution constraint, we find: 
\begin{small}
\begin{eqnarray}
P(\{\sigma\}^{\tau+1};\{\sigma'\}^{\tau}) & = & \frac{1}{Z(\{\sigma\})} \exp\left(\sum_{i=1}^{N}h_{i}^{\tau}\sigma_{i}'+\sum_{i,j=1}^{N}J_{ij}^{\tau}\sigma_{i}'\sigma_{j}'\right.\nonumber\\
&  & \left. +\sum_{i,j=1}^{N}J_{ij}^{\tau+1,\tau}\sigma_{i}\sigma_{j}'\right)
P(\{\sigma\}^{\tau+1})\label{eq:marginalization}
\end{eqnarray}
\end{small}
$Z(\{\sigma\})$ being the conditional partition function \Corr{, and $\{h_{i},J_{ij}\}_{i,j=1}^{N}$ are the Lagrange multipliers corresponding to the constraints given by $\{m_{i},C_{ij}\}_{i,j=1}^{N}$.}

We assume that the detailed balance is satisfied for a stationary distribution $P_{stat}(\{\sigma\})$. Therefore the Markovian matrix is also time-invariant
and satisfies the following relation
\begin{small}
\begin{equation}
P(\{\sigma'\}|\{\sigma\})P_{stat}(\{\sigma\})=P(\{\sigma\}|\{\sigma'\})P_{stat}(\{\sigma'\})
\end{equation}
\end{small}
so that:
\begin{small}
\begin{eqnarray}
& & P(\{\sigma\};\{\sigma'\})  =  P(\{\sigma'\}|\{\sigma\})P_{stat}(\{\sigma\})\label{eq:TransitionMatrix} \\
   & & =  \frac{\exp\left(\sum_{i=1}^{N}h_{i}\sigma_{i}+\sum_{i,j=1}^{N}J_{ij}\sigma_{i}\sigma_{j}+\sum_{i,j=1}^{N}J_{ij}^{1}\sigma_{i}'\sigma_{j}\right)}{Z(\{\sigma'\})}P_{stat}(\{\sigma'\}) \nonumber 
\end{eqnarray}
\end{small}

We then develop the extensive quantity $\ln\left(Z(\{\sigma'\})\right)$ up to the second order: 
\begin{small}
\begin{eqnarray}
\ln(Z(\{\sigma'\})) & = &	  \ln(Z_{eff})-\sum_{i=1}^{N}h_{i}^{r}\sigma_{i}'-\sum_{i,j=1}^{N}J_{ij}^{r}\sigma_{i}'\sigma_{j}'+\mathcal{O}(\delta\sigma'^{3})\label{eq:Zapprox}
\end{eqnarray}
\end{small}
The k-th order terms are k products of $J_{ij}^{1}$. This approximation is thus valid in the weak temporal correlation limit. 

Note that \Corr{the coefficients of this development,} $\{h_{i}^{r},J_{ij}^{r}\}_{i,j=1}^{N}$, can be obtained analytically from (\ref{eq:TransitionMatrix}) by a straightforward computation. 
The final form for the transition function then becomes: 
\begin{small}
\begin{eqnarray}
 &  & P(\{\sigma\}|\{\sigma'\}) = \frac{1}{Z_{eff}} \exp\left(\sum_{i=1}^{N}h_{i}\sigma_{i}+\sum_{i,j=1}^{N}J_{ij}\sigma_{i}\sigma_{j}+ \right.\nonumber\\
& & \left. \sum_{i,j=1}^{N}J_{ij}^{1}\sigma_{i}'\sigma_{j}+\sum_{i=1}^{N}h_{i}^{r}\sigma_{i}'+\sum_{i,j=1}^{N}J_{ij}^{r}\sigma_{i}'\sigma_{j}'\right)\label{eq:ApproxTransMat}
\end{eqnarray}
\end{small}

Using the detailed balance, the stationary distribution is then also restricted to the second order and has the generic form: 
\begin{small}
\begin{eqnarray}
P_{stat}(\{\sigma\}) & = & \frac{\exp\left(\sum_{i=1}^{N}h_{i}^{stat}\sigma_{i}+\sum_{i,j=1}^{N}J_{ij}^{stat}\sigma_{i}\sigma_{j}\right)}{\sum_{\{\sigma''\}}\exp\left(\sum_{i=1}^{N}h_{i}^{stat}\sigma_{i}''+\sum_{i,j=1}^{N}J_{ij}^{stat}\sigma_{i}''\sigma_{j}''\right)}\label{eq:Pstat}
\end{eqnarray}
\end{small}
Since 
\begin{small}
\begin{equation}
P_{stat}(\{\sigma\}) =\sum_{\{\sigma'\}} P(\{\sigma\}|\{\sigma'\})P_{stat}(\{\sigma'\})\label{eq:Stationnary}
\end{equation}
\end{small}
the parameters $\{h_{i}^{stat},J_{ij}^{stat}\}_{i,j=1}^{N}$
are fully determined by the $m_{i}$ and $C_{ij}$ values. 

Numerically, we adopt a slightly different approach, which is shown to be equivalent
to the approximation made above. 
We maximize separately the entropy of the stationary distribution $P_{stat}(\{\sigma\})$ and the time-invariant joint distribution $P(\{\sigma\};\{\sigma'\})$, without the marginalization condition. 
We obtain (\ref{eq:Pstat}) for $P_{stat}(\{\sigma\})$, and: 
\begin{small}
\begin{eqnarray}
 &  & P(\{\sigma\};\{\sigma'\}) = \frac{1}{Z_{tr}} \exp\left(\sum_{i=1}^{N}h_{i}\sigma_{i}+\sum_{i,j=1}^{N}J_{ij}\sigma_{i}\sigma_{j}+ \right.\nonumber\\
& & \left. \sum_{i,j=1}^{N}J_{ij}^{1}\sigma_{i}'\sigma_{j}+\sum_{i=1}^{N}h'_{i}\sigma_{i}'+\sum_{i,j=1}^{N}J'_{ij}\sigma_{i}'\sigma_{j}'\right)\label{eq:ApproxJointDistrib}
\end{eqnarray}
\end{small}

The transition matrix is then determined by: $P(\{\sigma\}|\{\sigma'\}) = \frac{P(\{\sigma\};\{\sigma'\})}{P_{stat}(\{\sigma'\})}$, which gives back (\ref{eq:ApproxTransMat}) if we identify $h_{i}^{r} = h'_{i} - h_{i}^{stat}$ and $J_{ij}^{r} = J'_{ij} - J_{ij}^{stat}$. 

This model contains seven sets of parameters, $\{h_{i},h_{i}^{stat},h_{i}^{r},J_{ij},J_{ij}^{stat},J_{ij}^{r},J_{ij}^{1}\}_{i,j=1}^{N}$.
In order to be equivalent to the previous model we must apply several
constraints which will reduce the number of free parameters. The stationary
parameters $\{h_{i}^{stat},J_{ij}^{stat}\}_{i,j=1}^{N}$ are bound
to the others by using the relation (\ref{eq:Stationnary}) as before.
Then we have to apply a normalization on the conditional probability
distribution (\ref{eq:ApproxTransMat}) to recover the marginalization
condition,  
which is a special form of (\ref{eq:Zapprox}) with $Z_{eff}=\frac{Z_{tr}}{Z_{stat}}$.
Therefore, the parameter set $\{h_{i}^{r},J_{ij}^{r}\}_{i,j=1}^{N}$
is also defined by $\{h_{i},J_{ij},J_{ij}^{1}\}_{i,j=1}^{N}$ which
are the only free parameters. This model is thus equivalent to the previous approximation
and allows for more tractable numerical treatements. 

To test the model, we \Corr{first used a} raster generated by a Glauber model\cite{Fischer1991}, \Corr{whose flip transition probability from one time step to the next is}
\begin{small}
\begin{eqnarray}
W(\sigma_{i} \rightarrow -\sigma_{i}) = \frac{1}{2\tau_{0}}\left(1 - \sigma_{i}(t) \tanh\left(\sum_{j} (J_{ij}^{g}\sigma_{j}(t) + h_{j}^{g}\sigma_{j}(t))\right)\right)\label{eq:Glauber}
\end{eqnarray}
\end{small}
where $\tau_{0}$ is the effective time constant and $J_{ij}$, $h_{i}$ are coupling constants of the neurons $\sigma$  \cite{note-Glauber-params}.

Fitting the model parameters to the corresponding $m_{\mathrm{i}}$, $C_{\mathrm{ij}}$ and $C_{\mathrm{ij}}^{1}$ values is a classical Boltzmann machine learning problem \cite{Ackley1985}. We started with an analytical approximation of the solution\cite{Tanaka1998} followed by a gradient descent: at each time step, the $m_{\mathrm{i}}$, $C_{\mathrm{ij}}$ and $C_{\mathrm{ij}}^{1}$ predicted by the model were estimated through a Monte-Carlo algorithm, compared to the experimental ones, and the model parameters were updated according to the difference. The algorithm was stopped when the difference between the theoretical and experimental values was less than 0.005, of the order of the uncertainty on the $m_{i}$ and $C_{ij}$ estimations.

In the following, we compared this model to simpler versions already used in the literature. The ``Ising model" has the same description of $P_{stat}(\{\sigma\})$, but assumed $P(\{\sigma\},\{\sigma'\}) = P_{stat}(\{\sigma\})P_{stat}(\{\sigma'\})$\cite{Schneidman2006,Tang2008} (this is equivalent to assume $C_{ij}^{1} = m_{i}m_{j}$). The ``independent model" assumed no second order interactions: all 
the previous parameters are null but the $h_{i}^{stat}$. 

\begin{figure}[!]
\centering
\vspace*{-0.6cm}
\hspace*{-0.7cm}
\includegraphics[width=0.6\textwidth]{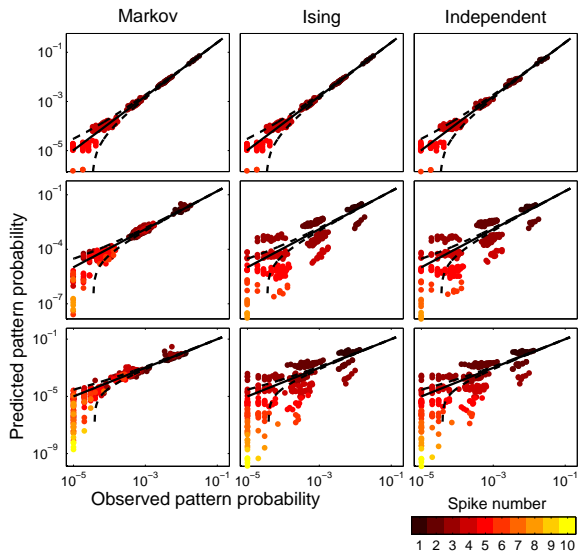}
\caption{(Color online) Performance of the 3 statistical models to describe the statistics generated by the Glauber model ($\tau_{0}=2$). For each panel, we compared the probability of several patterns estimated empirically from the raster, and predicted by the corresponding model. Each point corresponds to a different pattern, picked up in the raster. The point color indicates the number of spikes in each pattern. The black line indicates equality, and the dashed curves the 95\% confidence interval for the estimated probability. Each column corresponds to one of the three models described earlier. From left to right: the Markov, Ising and Independent models (see text). The different lines correspond to different pattern sizes (from top to bottom: 1, 2 and 3 temporal bins in the pattern).} 
\label{I2MFig9PredsGlauber}
\end{figure}

\begin{figure}[!]
\centering
\hspace*{-0.3cm}
\includegraphics[width=0.5\textwidth]{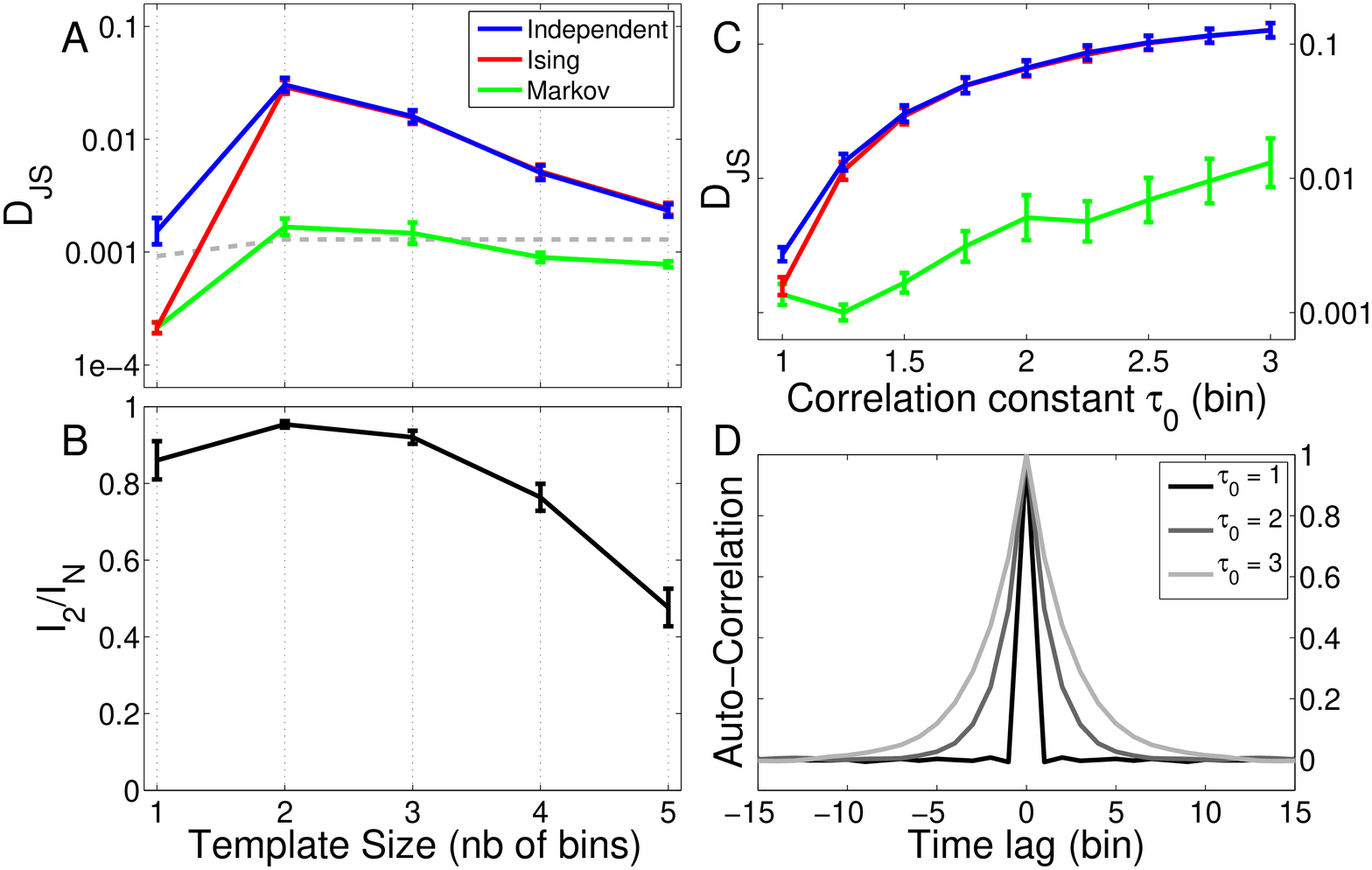}
\caption{(Color online) Quantification of the models performance. 
A: Jensen-Shannon Divergence $D_{JS}$ between the prediction of the three statistical models, and the probabilities estimated empirically, for different pattern sizes. The raster has been generated by the Glauber numerical model with parameter $\tau_{0}$=1.5. The gray line indicates the value below which $D_{JS}$ is not significantly different from zero (p $\le$ 0.01, \cite{Grosse2002}). 
B: Quantification with the information ratio $I_{2}/I_{N}$. 
C: Comparison for 2-bin pattern sizes, for different values of the $\tau_{0}$ parameter in the Glauber model.
\Corr{D: Auto-correlation of the population averaged activity for different $\tau_{0}$. }
} 
\label{I2MFigDjsEntGlauber}
\end{figure}

To estimate their performance in describing the statistics of the neural activity, we estimated the occurrence probability of several spiking patterns empirically and compared it to the ones predicted by each model. Figure \ref{I2MFig9PredsGlauber} shows the prediction of the three models for the probability of patterns with respectively 1, 2 and 3 time bins. For 
1-bin patterns, 
the Markov and the Ising model are equivalent, and showed a good prediction performance, 
with most of the points prediction being in the confidence interval of the estimated probability. 
For patterns with 2 and 3 time bins, the prediction remained satisfying for the Markov model, while it is strongly degraded for the Ising model. 
\Corr{Note that the Ising and independent models give similar performances here, contrary to \cite{Schneidman2006,Shlens2006}. Indeed, for a broad range of parameters in the Glauber model, the absolute correlation values are weak. However, their temporal extent controlled by $\tau_{0}$ (see Fig.~\ref{I2MFigDjsEntGlauber}D), is already sufficient to impair the Ising model performance.}

We quantified the fit between the model prediction and the experimentally measured statistics by computing the Jensen-Shannon Divergence: $D_{JS}(P,Q) = H(0.5(P+Q)) - 0.5(H(P)+H(Q))$ (where $H(\cdot)$ is the Shannon entropy) measures the similarity between two distributions P and Q \cite{Lin1991}. 
Figure \Corr{\ref{I2MFigDjsEntGlauber}A} shows the value of $D_{JS}$ for the three models, for different numbers of bins in the pattern. This confirmed our previous observation. For one bin, the Ising and the Markov model are equivalent, and performed better than the independent model. For two bins or more, the Markov model showed lower $D_{JS}$ values than the Ising model and the independent model. This prediction performance does not vary significantly with the number of bins. The Markov model is thus able to predict the probability of a pattern even when it is composed of several bins. \Corr{It thus describes with more accuracy} the statistics of the neural activity over a large temporal extent.

The better performance of the Markov model compared to the Ising model has to be related with the \Corr{shape} of the correlation functions: if the \Corr{temporal} correlation functions can be reduced to a \Corr{Dirac-like} form, there should be no difference between the Markov and Ising models \Corr{(case $\tau_{0} =1$ in Fig.~\ref{I2MFigDjsEntGlauber}C-D). 
Above 1, the normalized difference ${\delta}\log(D_{JS}) = (\log(D_{JS}^{Markov}) - \log(D_{JS}^{Ising})) / \log(D_{JS}^{Ising})$ quickly increases to reach a peak performance of 120\% around 2.5, and then slowly decreases to a plateau of 46 \% improvement from the Ising to the Markov model, for $\tau_{0} \ge 10$.
The Markov model thus performs better over a large range of $\tau_{0}$ values. From the experimental perspective, the Markov model prediction is at best when the ratio between the correlation time constant (Fig.~\ref{I2MFigDjsEntGlauber}D) and the bin size is around 2.5, but remains satisfying for larger ratios. }

We also computed the fraction of the ensemble correlations that was captured by the Markov model, $\frac{I_{2}}{I_{n}} = \frac{S_{1} - S_{2}}{S_{1} - S_{n}}$, where $S_{k}$ is the entropy when taking into account the correlations up to the k-th order \cite{Schneidman2003a,Schneidman2006}. 
This measures the improvement of the fit from the independent model to the Markov model. 
The value is maximal for two time bins, and then decreased (Fig.~\ref{I2MFigDjsEntGlauber}B), in line with the observed difference in $D_{JS}$ between the independent and the Markov model. This Markov model is thus able to explain a major part of the higher order spatio-temporal statistics. 

\begin{figure}[!]
\centering
\hspace*{-0.3cm}
\includegraphics[width=0.55\textwidth]{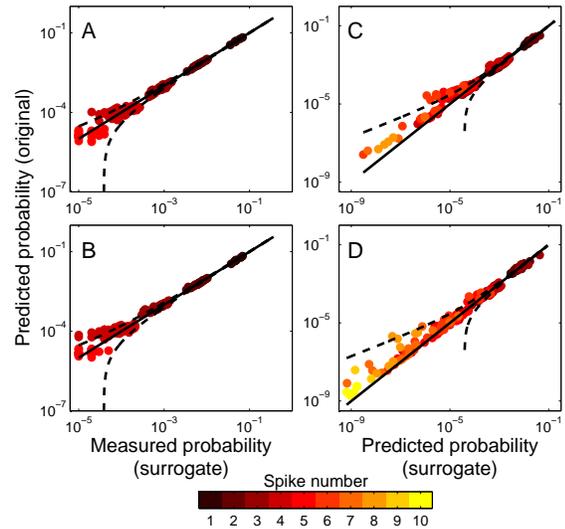}
\vspace*{-0.8cm}
\caption{(Color online) Tests of the surrogate raster generator. A: Comparison between the pattern probabilities in the surrogate raster, and the ones predicted by the model in the original analysis, for 1-bin patterns and a Glauber model with $\tau_{0}=1$ ($D_{JS}\simeq0.0003$). Same representation than in Fig.~\ref{I2MFig9PredsGlauber}.  
B: Same comparison than A for a Glauber model with $\tau_{0}=1.5$ ($D_{JS}\simeq0.0005$). 
C: Comparison between the prediction of the model fitted on the original data ($\tau_{0}=2$), and the prediction fitted on the surrogate raster, for 2-bins pattern ($D_{JS}\simeq0.0024$).
D: Same comparison than C for 3-bins patterns ($D_{JS}\simeq0.0024$).
} 
\label{I2MFigSurrogate}
\end{figure}

Apart from describing the statistics of the activity, this model can also be used to generate surrogate rasters having the same statistics than the captured ones. For that purpose, starting from an initial random pattern, we generate at each time step a new pattern according to (\ref{eq:ApproxTransMat}).
We then compared the statistics of this new raster with the original prediction (Fig.~\ref{I2MFigSurrogate}A). Although the generator only used the $h_{i}$, $J_{ij}$ and $J_{ij}^{1}$ coefficients of the model, the generated stationary probability is in very good agreement with the predicted stationary distribution estimated from the original data set, described by the $h_{i}^{stat}$ and $J_{ij}^{stat}$. This result shows the consistency of the model: 
the transition matrix defined by the $h_{i}$, $J_{ij}$ and $J_{ij}^{1}$ parameters has indeed the stationary distribution defined by the $h_{i}^{stat}$ and $J_{ij}^{stat}$ coefficients in (\ref{eq:Pstat}). 

We then applied the same analysis to the surrogate data, to obtain a model of the surrogate statistics. 
Fig \ref{I2MFigSurrogate}B shows that we recover the same predictions than with the original analysis. The generator is thus producing a surrogate raster \Corr{congruent} with the statistical model. 

\begin{figure}[!]
\centering
\hspace*{-0.3cm}
\includegraphics[width=0.5\textwidth]{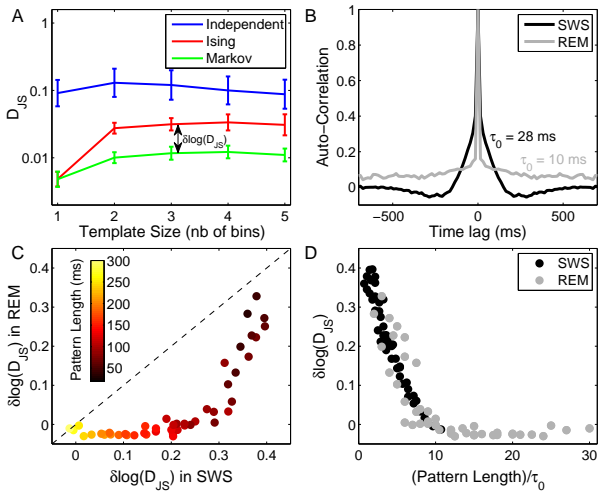}
\caption{
\Corr{
(Color online) Test of the models on experimental data. 
A: Jensen-Shannon divergence $D_{JS}$ for the 3 models, estimated for the activity of 8 channels in cat parietal cortex, and for different template sizes. Bin width of 10 ms.
B: Auto-correlation of the population averaged activity for the SWS and REM sleep states. The correlation time constants $\tau_{0}$ were estimated by fitting an exponential function. 
C: Relative log-difference ${\delta}\log(D_{JS})$ between the Markov and Ising $D_{JS}$, compared for the SWS and the REM data. The dotted line indicates equality. The different points correspond to different combinations of template and bin sizes, colour coded by the pattern length (template size x bin size). Points with black edge correspond to panel B values.
D: ${\delta}\log(D_{JS})$  for both states and for different pattern lengths, in unit of their respective correlation time constant (pattern length)/$\tau_{0}$. 
}
} 
\label{I2MFigData}
\end{figure}

\Corr{We then tested the model on {\it in vivo} biological data taken from \cite{Destexhe1999}, composed of 8 simultaneous multi-unit recordings in the cat parietal cortex in different sleep states (Slow Wave Sleep (SWS) and Rapid Eye Movement (REM)). For the activity recorded during SWS, the performance of the Markov model is significantly higher than for the Ising model. For a bin size of 10 ms, this was the case for different template sizes above 2 (Fig.~\ref{I2MFigData}A). The improvement was comparable to the difference between independent and Ising models. We estimated ${\delta}\log(D_{JS})$, the normalized log-difference between the Markov and Ising associated $D_{JS}$, for different combinations of template and bin sizes. The result holds, with $D_{JS}$ in the same order of magnitude, for larger bin sizes as long as the pattern length, defined as (template size) x (bin size), is below $\sim$120 ms (Fig.~\ref{I2MFigData}C). To see how the sleep state affects this result, we compared the ${\delta}\log(D_{JS})$ between the SWS and the REM activities (Fig.~\ref{I2MFigData}C). For pattern length below $\sim$120 ms, while the Markov model outperforms the Ising model in describing the SWS activity, the improvement drops rapidly for the REM state. For very large pattern lengths ($\sim$ 300 ms), the Markov and Ising models perform equally well (${\delta}\log(D_{JS})=0$) for both states. This faster drop of performance is related to the smaller correlation time constant in the REM state (Fig.~\ref{I2MFigData}B). This is indeed reminiscent of the case $\tau_{0}=1$ in the Glauber model (see Fig.~\ref{I2MFigDjsEntGlauber}C), and as a consequence, we observed no significant difference between the Ising and Markov models for intermediate pattern lengths. On the contrary, the SWS state exhibits larger correlation extent (similar to $\tau_{0} > 1$ in Fig.~\ref{I2MFigDjsEntGlauber}C), and shows a persistent difference ${\delta}\log(D_{JS})$. To futher emphasize this relation, we measured the correlation time constant $\tau_{0}$ for both states. We then computed ${\delta}\log(D_{JS})$ for different pattern lengths, expressed in unit numbers of their respective correlation time constant (pattern length)/$\tau_{0}$. When rescaled, both states exhibit the same dependency with the pattern length (Fig.~\ref{I2MFigData}D). The Markov model is thus suited for the analysis of temporally correlated activity for different data sets and for pattern lengths up to 10 times their correlation time constant.
}

In conclusion, we have presented a \Corr{probabilistic} model which gives an account of the distributed spiking activity with relatively few parameters, and takes into account both spatial and temporal pairwise correlations. 
The model still predicts the occurrence probability of larger temporal patterns, 
and can be used to generate surrogates which 
mimic the temporal and spatial correlation structure of the data. 
It would be interesting to test it on the \Corr{specific data that have been used to show the failure of the ising model \cite{Tang2008}.} 
Beyond spiking \Corr{assembly} activity, \Corr{other event-based data with long enough recordings might be interesting to
analyze} with this model \Corr{(for example calcium transients \cite{Stosiek2003})}. 
This method of analysis will help to tackle fundamental issues about the structure of the neural activity, like the existence of higher order statistics, or the Markovian nature of the temporal correlations. 
\Corr{It could also impact on a broad range of areas of physics and biology which used maximum entropy models} \cite{note-code}. 

We thank Michael Berry and Val\'erie Ego-Stengel for helpful
discussions. Experimental data were obtained with Diego Contreras
and Mircea Steriade, and were published previously \cite{Destexhe1999}. Support by CNRS, ANR (Natstats, HR-cortex), and EU (Bio-I3: Facets FP6-2004-IST-FETPI 15879) grants. O.M. was supported by DGA and FRM fellowships.


\begin{thebibliography}{99}
\bibitem{note-Glauber-params} In the following, we take a Glauber model of 8 units. The parameters $J_{ij}^{g}$ were uniformly chosen between -0.1 and 0.1, and the $h_{j}^{g}$ betwen -1.05 and -1.
\end{thebibliography}
\end{document}